# The impact of automation and connectivity on traffic flow and $CO_2$ emissions. A detailed microsimulation study.

*Michail Makridis*[1*], *Konstantinos Mattas*[2], *Biagio Ciuffo*[1] *and Georgios Fontaras*[1]

[1]European Commission Joint Research Centre, 21027 Ispra, Italy,
*michail.makridis@ec.europa.eu

[2]Democritus University of Thrace, Xanthi 67100, Greece

**Abstract**

The interest on the impact of vehicle automation and connectivity in the future road transport networks is very high, both from a research and a policy perspective. Results in the literature show that many of the anticipated advantages of connected and automated vehicles or automated vehicles without connectivity (CAVs and AVs respectively) on congestion and energy consumption are questionable. Some studies provide quantitative answers to the above questions through microsimulation but they systematically ignore the realistic simulation of vehicle dynamics, driver behaviour or instantaneous emissions estimates, mostly due to the overall increased complexity of the transport systems and the need for low computational demand on large-scale simulations. However, recent studies question the capability of common car-following models to produce realistic vehicle dynamics or driving behaviour, which directly impacts emissions estimations as well. This work presents a microsimulation study that contributes on the topic, using a scenario-based approach to give insights regarding the impact of CAVs and AVs on the evolution of emissions over a highway network. The motivation here is to answer whether the different driving behaviours produce significant differences in emissions during rush hours, and how significant is the impact of detailed vehicle dynamics simulation and instantaneous emissions in the outcome. The status of the network is assessed in terms of flow and speed. Furthermore, emissions are estimated using both the average-speed EMEP/EEA fuel consumption factors and a generic version of the European Commission's $CO_2$MPAS model that provides instantaneous fuel consumption estimates. The results of this work show that conservative driving of AVs can deteriorate the status of the network, and that connectivity is the key for improved traffic flow. Emissions-wise, the AVs have the highest fuel consumption per km travelled among other types, while CAVs only marginally lower the overall consumption of human-driven vehicles. For the same traffic demand, the total emissions for different vehicle types remain at comparable levels.

# Introduction

The impact of automation-related technologies on road transport networks is a topic studied for many years now (Kesting et al., 2010; Louwerse and Hoogendoorn, 2004). Connected and Automated vehicles (CAVs) are expected to bring significant advancements in the existing road transport systems in terms of reducing traffic congestion, energy consumption, $CO_2$ and pollutants emissions (Alonso Raposo et al., 2017; Litman, 2015).

The picture looks promising, but as research on the field progresses, many researchers express their doubts on the anticipated benefits, either energy or traffic-wise. In Fiori et al. (Fiori et al., 2019) preliminary results were presented showing that moving from internal combustion engine vehicles (ICEVs) to plug-in electric vehicles (PEVs), the relationship between congestion and energy consumption can change, with higher energy consumption connected to the free-flow test-cases. Mattas et al. (Mattas et al., 2018) presented a simulation case study on the impact of CAVs, focusing on the possible benefits of connectivity. The results show that depending on the traffic demand, the automated vehicles (AVs) can have adverse effects on traffic flow, while CAVs can be beneficial for the network, depending on their penetration rate.

At the same time, the tools used in various impact assessment studies have known limitations that may affect the final predictions. Results in the work of Ciuffo et al. (Ciuffo et al., 2018) raise concerns about the capability of the existing car-following models to reproduce observed vehicles' acceleration dynamics and thus estimate vehicles emissions and energy/fuel consumption. Most studies focus on congested conditions where the free-flow regime is expected to play a minor role. However, recent studies (Laval et al., 2014; Marczak et al., 2015) highlight that also in congested conditions, the acceleration regime, which incorporates vehicle dynamics and driving behaviour, affects the capacity drop, the hysteresis and possibly other traffic-related phenomena.

Panis et al.(Int Panis et al., 2006) highlighted the need for a detailed analysis of not only average speeds but also other aspects of vehicle operation such as acceleration and deceleration. On the same page, Lejri et al. (Lejri et al., 2018) proposed a model that accounts for traffic speed dynamics in order to provide more accurate emissions estimations. Finally, in the recent simulation study of Stogios et al. (Stogios et al., 2019), it is highlighted the importance of considering the different driving behaviors.

The prospect of CAVs in reducing the environmental impact of vehicles is of great importance. In this context, it is important to investigate how and to what extent CAVs technologies will affect vehicle energy use and reduce traffic emissions. On the other hand, if the technology does not deliver the expected results, it is essential to identify the correct traffic management strategies to help reach the desired emissions-reduction goals.

The present paper studies the impact of automation and connectivity on future road transport networks, based on certain assumptions regarding the vehicle technology, the traffic supply, and the demand. A microsimulation study on a highway network conducted using Aimsun traffic simulation software. The results are based on various scenarios with state of the art models for the simulation of different vehicle technologies involved, that is, conventional human-driven vehicles (CVs), AVs and CAVs. The same scenarios were replicated using the same models for the congested part but accounting explicitly the vehicle dynamics on free flow. Different technologies generate different driving behaviours that impact the levels of congestion in the network. The motivation of this work is to answer whether the variation in the driving behaviours produces significant differences in emissions during rush hours. Furthermore, it is important to understand the role of realistic vehicle dynamics simulation and how this impacts traffic flow and emissions.

The network used is the ring road of Antwerp presented in (Mattas et al., 2018). Initially, we build four scenarios using state of the art models for the simulation of the different vehicle types.. Each of the four scenarios refers to 3-hour simulation with a) human-driven vehicles, b) AVs, c) CAVs and d) CAVs with 20% increased traffic demand. Furthermore, in order to study whether the explicit simulation of vehicle dynamics can change these results, the above-mentioned car-following models were modified in order to explicitly consider realistic vehicle dynamics. In order to achieve the latter, we introduce the simulation of vehicle dynamics on free-flow, using the Microsimulation Free-flow aCceleration model (MFC) (Makridis et al., 2019a). The above-mentioned four scenarios are replicated using now the modified car-following models, leading to 8 scenarios in total for this work. Results demonstrate the impact of connectivity and automation on traffic flow and emissions.

The reference models used for the assessment of $CO_2$ emissions estimates are the fuel consumption and $CO_2$ emissions factors proposed by the EMEP/EEA guidebook, and a generic version of $CO_2$MPAS (Fontaras et al., 2018) similar to the one described in (Tsiakmakis et al., 2017). The EMEP/EEA guidebook methodology (European Environment Agency, 2016), is more widely known by its software implementation, COPERT. It foresees an average-speed model and is frequently used in most European countries in order to estimate emissions of all major air pollutants produced by different vehicle categories. $CO_2$MPAS is a vehicle-specific simulation model recently introduced by the EU in its vehicle $CO_2$ certification system (Fontaras et al., 2018). Finally, it should be highlighted that the results of this study are bounded by a set of assumptions, which are summarized a) in the accuracy of the car-following models used for the simulation of different vehicle types, b) the accuracy of the emissions models, c) the accuracy of the vehicle dynamics simulation model and d) the absence of electric vehicle dynamics simulation.

**Simulations**

This section describes the driver models used in order to simulate the different technologies, the models used to provide emissions estimations and finally, the physical network and the different scenarios.

**Driver models**

AIMSUN traffic simulation software is used in this study. Six combinations of driver and vehicle types were simulated. The first three are state of the art car-following models that have been regularly used for the simulation of human-driven, automated, and connected and automated vehicles (CVs, AVs and CAVs respectively). All of these models are separated into two main parts, the free-flow and the congested one. As discussed in the work of Ciuffo et al. (Ciuffo et al., 2018) existing car-following models cannot reproduce with high accuracy vehicle acceleration dynamics as shown in empirical observations and thus the provide low accuracy estimates regarding vehicles emissions and energy/fuel consumption. Consequently, in order to simulate in detail the dynamics of the vehicle and see its impact on traffic flow and emissions, each of the car-following models mentioned above was coupled with the Microsimulation Free-flow aCeleration model (MFC) (Makridis et al., 2019a), a dynamics-based model which also incorporates the driver behaviour. More specifically, the MFC model was used to represent the uncongested part of the three above-mentioned models leading to three additional models that take into account the vehicle dynamics in an explicit way. The three new hybrid models derive from the original formulas by substituting their free-flow acceleration parts $a_{ff}$, with the corresponding one from the MFC model. It should be noted that this work does not use a distribution of vehicles. Vehicle dynamics simulation used is based on the technical specifications of a vehicle which is considered typical of the European segment C for passenger vehicles. Regarding variability in the driving behaviors, a uniform distribution of drivers is used, where the parameters of the MFC model $\{DS, GS_{th}\}$ take values between 0.4 and 0.9, where 0.4 corresponds to a timid driver and 0.9 to an aggressive one. The same driving behaviors were used also for AVs and CAVs assuming that the behaviour of the automated driving control systems will mimic the human-driven vehicles as it is mentioned in (Makridis et al., 2019b). Regarding overtaking, lane changing and giving way behaviours, the default AIMSUN models have been used, taking into account the different reaction times and deceleration capabilities for AVs and CAVs. It should be noted that in the rest of the document the term reaction time is used for human-driven vehicles, i.e., CVs and signifies that delay between the action of the leading driver and the reaction from the following driver. The corresponding delay for AVs and CAVs, is called controller's response time.

***Conventional Vehicles (CVs):*** In this study, conventional vehicles are considered those that are human-driven, have no automated functionalities and no connectivity. For the simulation of human-driven vehicles, the default model that is implemented in AIMSUN was used. This is a modified Gipps' car-following model (Gipps, 1981). The Gipps model can be described as follows:

$$v_n(t+\tau) = \min\{v_n(t) + 2.5 * a_n\tau \left(1 - \frac{v_n(t)}{V_n}\right)\left(0.025 + \frac{v_n(t)}{V_n}\right)^{\frac{1}{2}}, b_n\tau + \sqrt{(b_n^2\tau^2 - b_n[2[x_{n-1}(t) - s_{n-1} - x_n(t)] - v_n(t)\tau - v_{n-1}(t)^2/\hat{b}]]}\} \quad (1)$$

Where $a_n$ is the maximum acceleration, $b_n$ is the most severe braking, $s_n$ is the effective size of the vehicle, $V_n$ is the desired speed, $x_n(t)\ and\ v_n(t)$ are the location and speed of the vehicle $n$ at time $t$, and $\tau$ is the reaction time. The model can be expressed in a more abstract way as:

$$v_{n,t+\tau} = v_{n,t} + \min\{a_{ff}, a_{cf}\} * t \quad (2)$$

Where $\tau$ is the reaction time, $a_{ff}$ the free-flow acceleration and $a_{cong}$ the acceleration under car-following.

***Automated Vehicles (AVs):*** In this study, automated vehicles are considered those that are completely autonomous without any human intervention but have no connectivity capabilities. For the simulation of the AVs, the model proposed in (Shladover et al., 2012) is used. It is based on the ACC car-following rules that are proprietary to Nissan and to the best of the authors' knowledge, it is considered one of the most representative models regarding the microsimulation of automation while car-following. It is a first-order model representing ACC vehicle longitudinal behaviour. For the lateral movement, the default AIMSUN model was used, according to the ACC maximum deceleration and car following deceleration functions. The acceleration under car-following for the model is described by the following equation:

$$a_{cf} = k_1(d - t_w v_{n-1}) + k_2(v_{n-1} - v_n) \qquad (3)$$

Where $k_1 \: and \: k_2$ are constants, $t_w$ is the desired time-gap and $d$ is the current inter-vehicle distance. For more details, please refer to (Shladover et al., 2012).

The acceleration under free-flow is described as follows:
$$a_{ff} = \max(\min(k(V_n - v_n), a_n), b_n) \qquad (4)$$

Where $k$ is a constant, $a_n$ is the maximum acceleration, $b_n$ is the maximum deceleration, $v_n$ is the speed of the follower and $V_n$ is the desired speed of the follower. More abstractly we have:

$$v_{n,t+\tau} = v_{n,t} + \min\{a_{ff}, a_{cf}\} * t \qquad (5)$$

Where, $\tau$ is the response time, $a_{ff}$ the free-flow acceleration and $a_{cf}$ is the acceleration under car-following.

***Connected Automated Vehicles (CAVs):*** In this study, connected and automated vehicles are considered those that are completely autonomous and they have connectivity capabilities. In order to simulate the longitudinal movement of CAVs, we use the model described by (Talebpour and Mahmassani, 2016), which is an extended version of the Cooperative-ACC model proposed by (Van Arem et al., 2006). The extended version used here adds a set of constraints that practically check if a leader is spotted, and in that case it ensures that the speed of the autonomous vehicle is low enough to allow it to stop if its leader decides to decelerate with its maximum deceleration rate and reach a full stop. CAVs are forced to obey the speed limits, in the same way as AVs. Lane changing is again modelled based on the default AIMSUN algorithm, using the CAVs particular car following deceleration model.

$$a_{ff} = k(V_n - v_n) \qquad (6)$$

Where $V_n$ is the desired speed and $v_n$ the current speed of the follower.

$$a_{cf} = k_a a_p + k_v(v_{n-1} - v_n) + k_d(r - r_{ref}) \qquad (7)$$

Where $k_a, k_v,$ and $k_d$ are constant factors, $v_{n-1}$ and $v_n$ the speed of the leader and the follower, $a_p$ the acceleration of the leader, $r$ and $r_{ref}$ denote the current and reference clearance to the leading vehicle. For implementation details, please refer to (Talebpour and Mahmassani, 2016).

Finally, the model can be described in the same formula as above:

$$v_{n,t+\tau} = v_{n,t} + \min\{a_{ff}, a_{cf}\} * t \qquad (8)$$

Where, $\tau$ is the reaction time, $a_{ff}$ the free-flow acceleration and $a_{cf}$ is the acceleration under car-following.

***MFC free-flow model:*** In this study, the vehicle dynamics simulation is performed only on free-flow by the MFC model. The Microsimulation Free-flow aCeleration model is a vehicle dynamics-based model for the estimation of the vehicle's free-flow acceleration based on the specifications of the vehicle and the driving style of the driver.

$$a_{ff} = a_w \cdot a_{cp} \qquad (9)$$

Where $a_w$ is the willingness of the driver to accelerate and it is defined according to the driver's driving style and $a_{cp}$ is the acceleration potential of the vehicle based on its technical characteristics.

MFC is able to capture the vehicle acceleration dynamics accurately and consistently, it provides a link between the model and the driver and can be easily implemented and tested without raising the computational complexity. The proposed model is calibrated, validated and compared with known car-following models on road data on a fixed route inside the Joint Research Centre of the European Commission. Finally, the MFC has been validated based on 0-100km/h acceleration specs of thousands of vehicles available in the market. The results show the robustness and flexibility of the model. Implementation details can be found in (Makridis et al., 2019a).

As mentioned above, this study does not use a distribution of vehicles or drivers. The MFC used the technical specifications of a representative commercial European segment C passenger vehicle. In order to simulate variability in the driving behaviour, a uniform distribution of drivers is used, where the parameters of the MFC model $\{DS, GS_{th}\}$ take values between 0.4 and 0.9, where 0.4 corresponds to a timid driver and 0.9 to an aggressive one.

Table 1 presents the main model parameters. The maximum parameter values reported in the table are used in AIMSUN as the means of normal distributions in the drivers' population. AVs and CAVs have stricter limits on accelerations and decelerations. These are aimed to increase passengers' comfort. Furthermore, the term reaction time for human-driven vehicles signifies that delay between the action of the leading driver and the reaction from the following driver. For AVs and CAVs, this time is called the controller response time. Response times are also much lower for AVs and CAVs than reaction times of CVs. However, they are not negligible, as this would contradict with empirical evidence (Makridis et al., 2018). For other driving parameters, not shown in Table 1, the default values in AIMSUN were used for human-driven vehicles. The rest of the parameter values within the CV, MFC, AV and CAV models are those reported in the corresponding references.

**Table 1** Main model parameters

| Property/Vehicle type | CVs model | AVs model | CAVs model |
|---|---|---|---|
| Vehicle length (variation) | 4m (0.5m) | 4m (0.5m) | 4m (0.5m) |
| Vehicle width (variation) | 2m (0m) | 2m (0m) | 2m (0m) |
| Reaction (response) time | 0.8s | 0.3s | 0.3s |
| Time gap | n/a | 1.6s | n/a |
| Max. acceleration | $3\ m/s^2$ | $2\ m/s^2$ | $2\ m/s^2$ |
| Max. deceleration | $-6\ m/s^2$ | $-3\ m/s^2$ | $-3\ m/s^2$ |

**Predicting models for $CO_2$ emissions**

The two models chosen for this exercise are $CO_2$MPAS and EMEP-EEA guidebook (EMEP). The EMEP methodology is based on an average-speed model, frequently used in most European countries in order to estimate emissions from all major air pollutants produced by different vehicle categories. It is an implementation of the emissions inventory guidebook (European Environment Agency, 2016), the reference instrument designed to facilitate reporting of on-road emissions in European countries, allowing for a transparent and standardised, hence consistent and comparable, emissions reporting procedure.

As mentioned in the introduction, $CO_2$MPAS is a vehicle $CO_2$ emissions and energy consumption simulator, created for the introduction of the WLTP test protocol in the European legislation. The core of $CO_2$MPAS is a longitudinal dynamics physical model simulating energy flow and losses at various components. It operates using as input information regarding the vehicle (e.g. mass, road loads, tyre type etc), components (e.g. gearbox type, ratios, number of gears), and the engine (e.g. max power, capacity, maximum torque output) (Fontaras et al., 2018). $CO_2$MPAS is an open-source tool, and it is available online with all its documentation(European Commission, 2015). In this work, we use a generic version ($CO_2$MPAS-generic) operating in a mixed forward/backwards implementation, with the MFC driver model defining the acceleration achieved at each time-step based on the network conditions, and the $CO_2$MPAS generic calculating the energy flows in each vehicle following a similar approach as presented by (Tsiakmakis et al., 2017). The $CO_2$MPAS generic is tested against the EMEP/EEA methodology and shown to produce better and unbiased estimations (Mogno et al., 2019).

**Network and scenarios**

The case study network for the simulation experiments is the ring road around Antwerp, Belgium, as depicted in Figure 1. The ring road's specifications were extracted from Open Street Maps and refined, resulting to a network consisting of 119km of roads with 27 centroids (origin/destination points), 208 sections with variable numbers of lanes and 117 intersections. There are no traffic lights on the network. Due to the ring road shape, there are obvious paths for each O/D pair, so no distinction has been made between user equilibrium and system optimum, although this can amplify the benefits of connectivity in different situations.

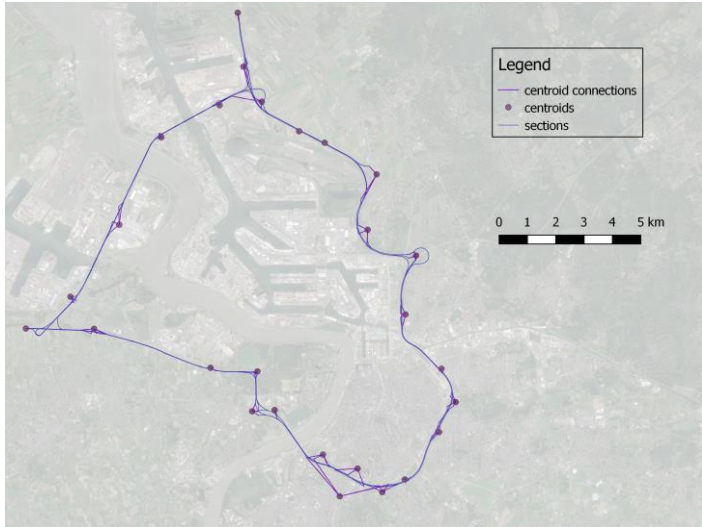

**Figure 1:** The ring road model of Antwerp, Belgium

Traffic count data during the morning peak hour were utilised to produce the base scenario traffic demand. The Frank and Wolfe algorithm (Frank and Wolfe, 1956), which is available as a built-in tool in AIMSUN was used to adjust a planning O/D matrix to the observed data.

Eight scenarios have been designed to support this study on the ring road of Antwerp:

- Scenario 1: Only CVs.
- Scenario 2: AVs (no V2V communication).
- Scenario 3: CAVs (V2V communication).
- Scenario 4: CAVs (V2V communication) and increased traffic demand by 20%
- Scenario 5: CVs using the MFC for the simulation of the free-flow acceleration.
- Scenario 6: AVs using the MFC for the simulation of the free-flow acceleration.
- Scenario 7: CAVs using the MFC for the simulation of the free-flow acceleration.
- Scenario 8: CAVs using the MFC for the simulation of the free-flow acceleration; moreover, increased traffic demand by 20%

All scenarios assume full penetration rate of the corresponding vehicle type (CVs, AVs, CAVs).

Different technologies generate different driving behaviours that impact the network's overall congestion levels. Each basic scenario was run twice; using the original free-flow acceleration of the model (Scenarios 1-4) and using the proposed free-flow acceleration by the MFC vehicle dynamics-based model for an average vehicle of segment C and a uniform distribution of different driving behaviours according to the MFC driver parameters (Scenarios 5-8) . Each scenario refers to three hours of simulation with the second hour being the network's peak hour. The first and the last hours are loading and unloading periods with lower demand equal to 20% of the observed peak-hour demand. All different scenarios were simulated with the same random seed number for efficient comparison between different scenarios.

It should be noted that instantaneous emissions estimations cannot be produced for the traditional car-following models such as the ones used in Scenarios 1-4 for the simulation of the different vehicle types. $CO_2$MPAS generic provides instantaneous emissions estimations only if detailed vehicle-related information, such as vehicle mass, power, gear ratio, gear in the box per simulation step, clutch condition (pressed or released) etc. is given as input. Such information is available only when the MFC model is used. Furthermore, it has been found that microsimulation models in some cases (e.g. higher speeds or after severe braking) produce unrealistic acceleration values (and thus questionable emissions estimates), above the realistic vehicle power potential (Ciuffo et al., 2018). Consequently, $CO_2$MPAS generic estimations are available only for Scenarios 5-8. On the other hand, results from EMEP/EEA methodology, which is based on the average speeds, are presented for all 8 scenarios.

## Results

Simulated data regarding the state of the network were retrieved for 10-minute intervals. The results focus on three major dimensions, the impact of the vehicle/driver technology, i.e. CVs, AVs or CAVs, the impact of the simulation of vehicle dynamics, i.e. MFC and finally, the differences in the emissions estimations from EMEP/EEA and the instantaneous generic $CO_2$MPAS.

Figure 2 illustrates the evolution of the average harmonic speed per 10-minute period over the 3-hour simulation for different vehicle types and all scenarios. During the first hour, the network is loading and the demand is quite low. Consequently, all vehicle types manage to maintain high average speeds close to the road speed limits. Since CVs (Scenarios 1 and 5) do not always obey the speed limit, this behaviour is reflected in their average harmonic speed as well.

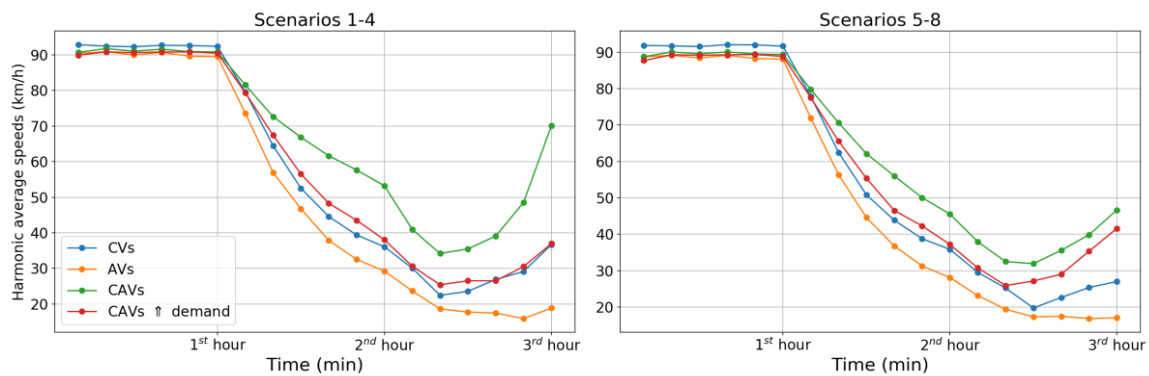

**Figure 2:** Harmonic average speed for CVs, AVs, CAVs and CAVs with increased demand for the eight different scenarios. On the left, the vehicle types do not include vehicle dynamics simulation. On the right, the MFC model was applied.

On the end of the first hour, the demand is increasing to its peak value, and the network starts to become saturated. When the vehicle dynamics are not simulated (Scenarios 1-4), the vehicles accelerate sharper and consequently, they are able to maintain higher speeds and also offload the network quicker after the second hour. Vehicle dynamics simulation (Scenarios 5-8) leads to lower average speeds. Furthermore, increased demand (Scenarios 4 and 8) leads to the worst results in terms of speed. As expected since CAVs (Scenarios 3 and 7) have the ability for Vehicle-to-Vehicle (V2V) communication, they can keep shorter time headways, have average speeds that are significantly higher than the other two vehicle types (CVs and AVs) and maintain a better status for the network.

As it is shown in Figure 2, the introduction of vehicle dynamics leads to more conservative accelerations on the road. More specifically, for the same scenarios, the explicit simulation of vehicle dynamics reflects to a delay in the unloading of the network. During the 3rd (network's unloading phase), the average speeds in Scenarios 1-3 start to increase earlier than in Scenarios 5-7. The observations in this study are aligned with the conclusions in (Makridis et al., 2019a) that traditional car-following models used in simulation software tend to be overly aggressive, leading, at time, to sharp accelerations that do not correspond to realistic vehicle power potential. This behaviour in some cases prevents the network from becoming saturated. However, this is an artefact as standard vehicles do not have the capability to accelerate so sharply.

Additionally, it seems that automation alone (Scenarios 2 and 6) is not capable of improving the situation of the transport network. On the contrary, the driving behaviour of the AVs is more conservative with less sharp acceleration and desired speed fixed to the road limits, resulting in a more saturated network for the same demand.

Figure 3 depicts the results in terms of network outflow, which is the number of vehicles per hour that exit the network for each ten minute interval. It is worth noting that AVs (Scenarios 2 and 6) are more conservative leading to the lowest throughput when the network becomes congested. It is interesting to note that after the simulation run, the AVs do not manage to serve the requested demand like CVs and CAVs (some vehicles are still in the network after the 3rd hour of the simulation). The CAVs with increased demand (Scenarios 4 and 8), as expected, have a higher throughput, which is considered very close to the network's capacity. However, even with the normal peak demand (Scenarios 3 and 7), CAVs perform very well, approaching the capacity of the network, while for CVs (Scenarios 1 and 5) but most notably for AVs (Scenarios 2 and 6), the average throughput decreases a lot.

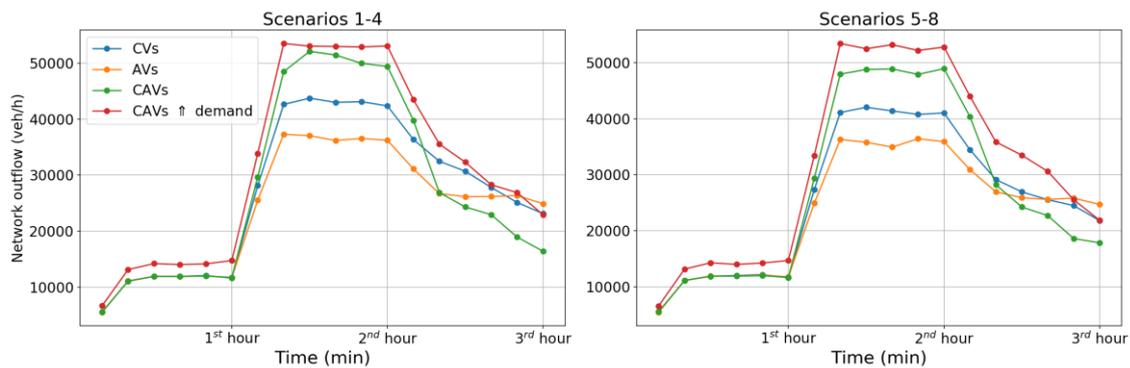

**Figure 3:** The number of vehicles per hour that exit the network is shown in this figure for the different vehicle type, i.e. for CVs, AVs, CAVs and CAVs with increased demand. The figure on the left shows the results for Scenarios 1-4, while the figure on the right shows the results for Scenarios 5-8 (models with free-flow vehicle dynamics simulation).

**Table 2** Mean speed and total flow per scenario

| Scenarios | Mean speed ($km/h$) | Flow ($veh/h$) |
| --- | --- | --- |
| Scenario 1 | 59.7 | 26817 |
| Scenario 2 | 52.9 | 24133 |
| Scenario 3 | 71.5 | 27472 |
| Scenario 4 | 62.6 | 31428 |
| Scenario 5 | 59.2 | 25567 |
| Scenario 6 | 51.9 | 23786 |
| Scenario 7 | 67.5 | 27098 |
| Scenario 8 | 61.3 | 31431 |

Table 2 summarizes the macroscopic indicators from the network simulation per simulation scenario. From the table, it is clear that the vehicle dynamics do not have a strong impact on the overall picture in terms of mean speed and flow. Scenarios 1 and 5, 2 and 6, 3 and 7 as well as 4 and 8 have very similar mean speed and flow values. From a vehicle technology point of view, it is important to mention that although the behaviour of each vehicle type is different, this is not depicted in the global traffic indicators. Therefore, any assessment of the AVs or CAVs behavior should be performed at a micro-level. From an emissions point of view, it is normal to expect that the EMEP/EEA methodology which is based on the average speeds will give similar results for Scenarios 1-4 with Scenarios 5-8. Consequently, it is reasonable to conclude that the differences in the emissions results between the first and last four scenarios are mainly due to the different vehicle dynamics and the application of the MFC model. Finally, it is also clear that an accurate vehicle dynamics model can contribute significantly to the estimation of emissions and highlight clearer the differences between the behaviours of the different vehicle technologies.

**Local $CO_2$ emissions predictions**

Any microsimulation study incorporates a great amount of stochasticity due to the different models, the network, the distributions of drivers or vehicle specifications, the sequence that vehicles enter the network etc. Consequently, the production of highly precise quantified results is not an easy task. On the other hand, many models are nowadays calibrated and validated with high precision empirical observations. Consequently, a comparison study between the results of different models can at least point to the correct direction. The proposed study leads to very interesting relative comparisons and provides significant qualitative conclusions. For example, the impact of vehicle dynamics simulation can be directly inferred if we compare Scenarios 5-8 with Scenarios 1-4. On the same page, the impact of instantaneous emissions estimates can again be observed if we compare the output of both emissions models for the Scenarios 5-8.

The figures below present the change in the total fuel consumption (and consequently $CO_2$ emissions that are directly proportional to fuel consumption) compared to those of the basic scenario of 100% conventional vehicles in the fleet. Many simulation studies provide fuel consumption estimations based on the EMEP/EEA methodology, which foresees an average speed model. However, vehicle dynamics affect a lot the instantaneous fuel consumption of each vehicle, and it is interesting to see whether this instantaneous deviation is reflected in the overall results over a network. We use a generic version of $CO_2$MPAS to provide fuel consumption estimations for the scenarios when the MFC model is used to simulate the vehicle dynamics (Scenarios 5-8). This model cannot be used for Scenarios 1-4, as it is dependent on the vehicle dynamics, which are unrealistic without the use of the MFC model (extreme accelerations observed under the standard model).

Figure 4 shows the efficiency of the different vehicle types in emissions per kilometer and in comparison with CVs. The green color corresponds to lower emissions, while the red corresponds to higher one. Figures 4a, b and c present the results for the Scenarios 1-4 and the EMEP/EEA methodology, the Scenarios 5-8 and the EMEP/EEA methodology and the Scenarios 5-8 and the $CO_2$MPAS methodology. The results presented here are a percentage relative to the estimated emissions of CVs for the corresponding scenarios.

Fig. 4a and Fig. 4b illustrate that according to EMEP/EEA model, CAVs produce fewer emissions per kilometer than CVs in total while results from the $CO_2$MPAS generic model (Fig. 4c) reverse this observation by a small margin from -2% to 1% more. This is quite significant, as it changes the final qualitative conclusion on which vehicle type produces higher emissions. Furthermore, in all cases, the AVs (Scenarios 2 and 6) seem to exhibit the highest consumption than the other vehicle-types, even in the case of CAVs with increased traffic demand. As mentioned in the relevant literature (Mattas et al., 2018), automation alone is not expected to improve the situation in current transport networks, while connectivity seems very important.

Using realistic vehicle dynamics changes the overall emissions even for the EMEP/EEA methodology. It is interesting to note that the efficiency gap between AVs and CAVs (difference on Scenarios 2-3, Fig. 4a versus the difference on Scenarios 6-7, Fig.4b) becomes smaller. Moreover, this gap becomes even smaller, almost insignificant when the estimations are provided by the generic $CO_2$MPAS (Fig. 4c). Another interesting observation is the CAVs with increased demand and 20% more vehicles in the network produce the same emissions as AVs with standard peak demand (Fig. 4c). In all scenarios, the differences in emissions between CVs, AVs, CAVs and CAVs with increased demand range from -3.4% to 6% for the most extreme cases.

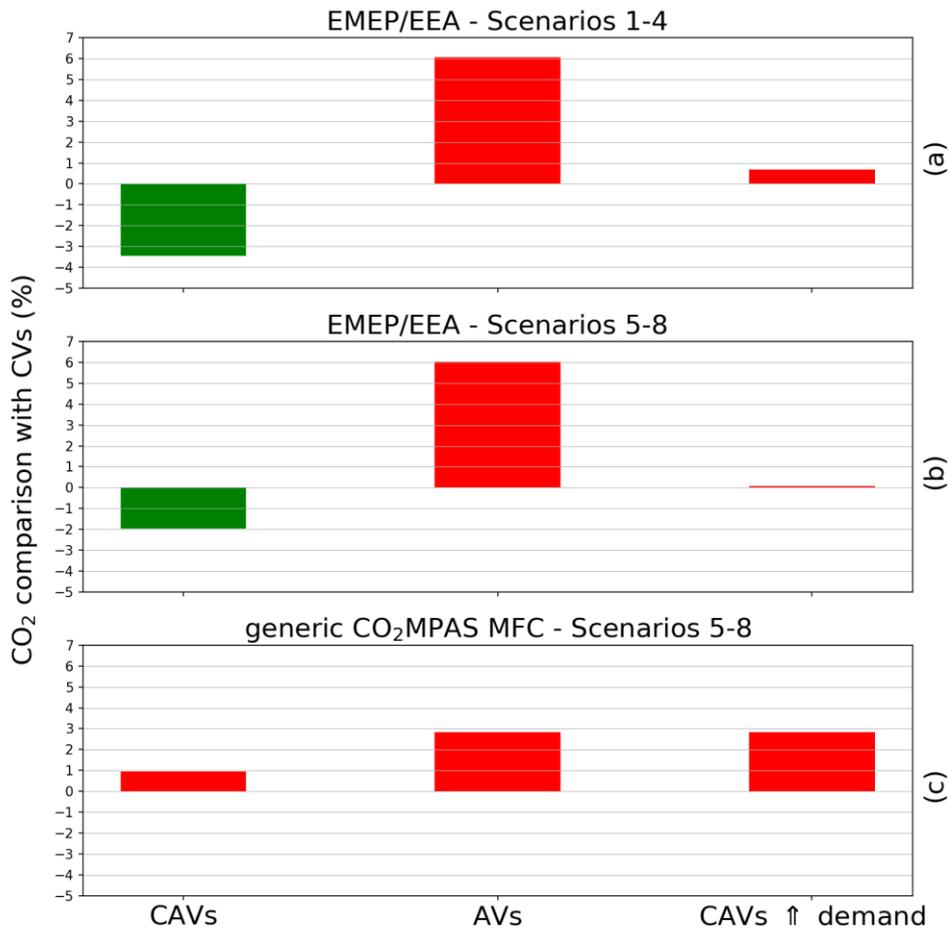

**Figure 4:** The total fuel consumption of AVs, CAVs and CAVs with increased traffic demand, in comparison with the corresponding consumption of CVs. The results presented here are a percentage relative to the estimated emissions of CVs for the corresponding scenarios.

Finally, it is interesting to observe the evolution of the total emissions produced per scenario. Figure 5 illustrates the comparative results for 10-minutes intervals during each 3-hour simualation. The reference number in this figure is the total emissions produced by CVs. Fig. 5a-c present the results for the Scenarios 1-4 and the EMEP/EEA methodology, the Scenarios 5-8 and the EMEP/EEA methodology and the Scenarios 5-8 and the $CO_2$MPAS methodology. As expected, the CAVs with increased demand (Scenarios 4 and 8) by 20% produce much higher consumptions, yet it is interesting to note that the difference is not linearly correlated with the increase in the demand. When the network is empty, during the first hour, the increase in the demand is translated to the same increase in emissions (~20%). However, during the peak hour (2nd hour), the increase in fuel consumption seems to be related to the capacity of the network. Since more and more CAVs get into the network, the increase in the emissions reaches almost 35%. Correspondingly, during the last hour that the network unloads, most of the demand has served and consequently, the consumption drops to the values of CVs in scenario 1. It is interesting to notice that in Fig.4 the emissions per kilometer are presented for each different vehicle type. AVs (Scenarios 2 and 6) produce less aggressive driving behaviour, which would imply better efficiency. However, AVs have a negative impact on the traffic flow, which eventually leeds to higher emissions per kilometer. On the other hand, in Fig.5 the presented figures are in total emissions. Having already presented estimations of vehicle efficiency, the total emission rates are of interest. Moreover, since the figure presents 10 minute intervals, the rate of emitting is estimated. It is shown that with increased capacity, more vehicles fit in the network, travelling with higher speeds. Hence, as presented, the rate of emitting can be increased by more than 30% comparing to contemporary traffic. On the other hand, the rate of emissions for AVs is smaller during congestion, which is a result of less vehicle kilometers driven.

The simulation of vehicle dynamics amplifies the differences between different types of vehicles. For example, the CAVs with the generic $CO_2$MPAS model produce 20% more fuel consumption during the peak hour and 20% less during the last hour. This difference is not visible in the results of EMEP/EEA for Scenarios 1-4, starts to become visible in the results of EMEP/EEA for Scenarios 5-8 and it's clearly observable in the instantaneous estimations.

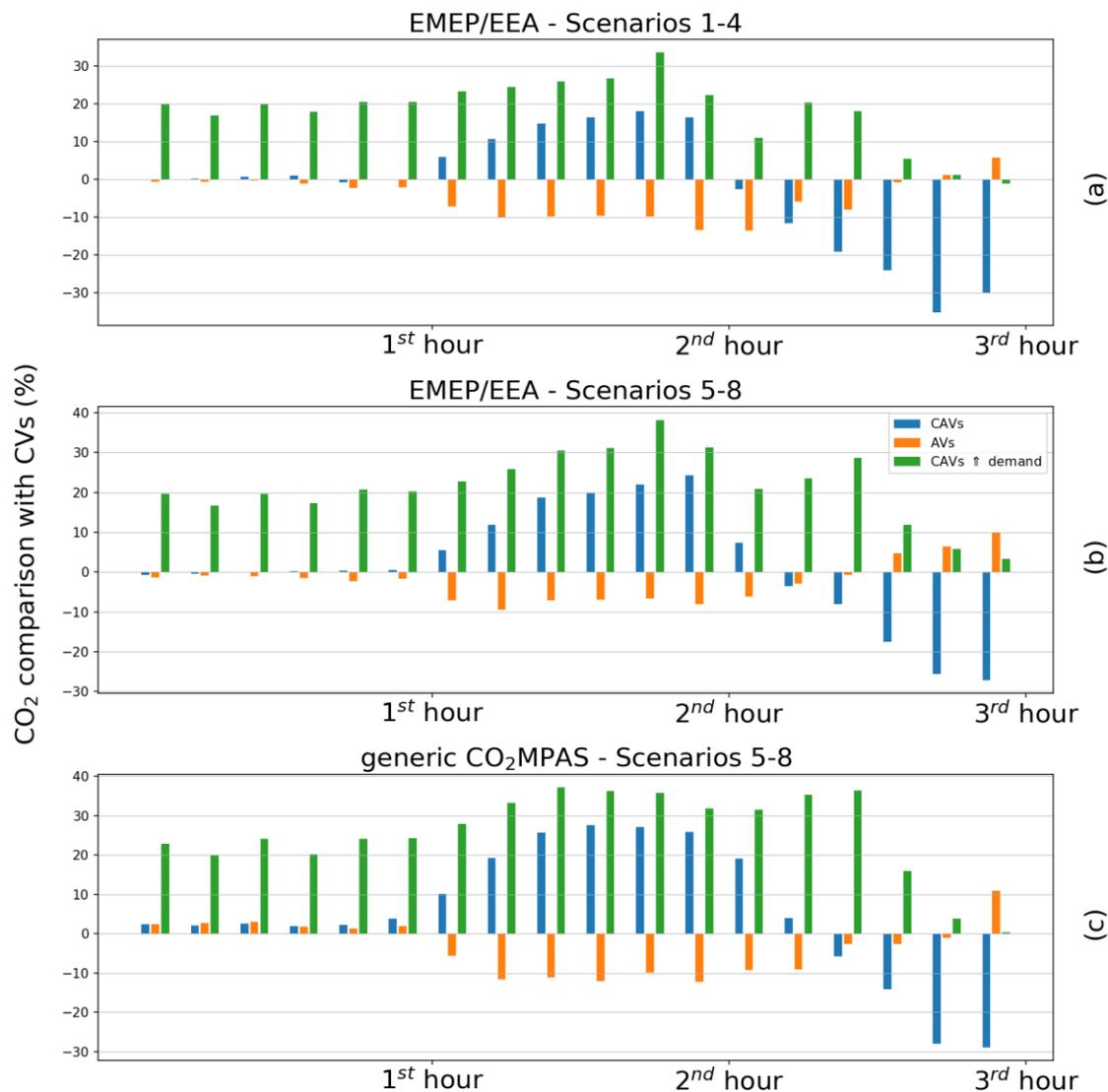

**Figure 5:** Fuel consumption estimations per 10-minutes periods for AVs, CAVs and CAVs with increased traffic demand, in comparison with the corresponding consumption of CVs. The results presented here are a percentage relative to the estimated emissions of CVs for the corresponding scenarios.

## Discussion and conclusions

This work proposes a microsimulation framework that investigates the impact of vehicle automation and connectivity in terms of traffic flow and emissions on a realistic highway transport network, the ring road of Antwerp, Belgium. The proposed framework takes into consideration by modelling the different vehicle technologies that generate different vehicle behaviours and traffic patterns. Moreover, it investigates the potential impact of simulating realistic vehicles dynamics through modelling. The study was conducted using the Aimsun traffic simulation software. Four different scenarios are tested to compare human-driven vehicles (CVs), automated vehicles (AVs), connected and automated vehicles (CAVs) and finally CAVs in a scenario with 20% increased traffic demand. Each vehicle type is simulated with the corresponding model for CVs, AVs and CAVs, models that are considered state of art in the literature. Especially the models for AVs and CAVs are calibrated on empirical observations. Next, additional four scenarios are generated using the same models for car-following but the free-flow terms are imposed by a vehicle dynamics-based model. More specifically, the models are combined with the Microsimulation Free-flow aCceleration model (MFC) that simulates the vehicle dynamics and the driving style explicitly. Each of the eight scenarios lasts 3 hours, one hour for network loading, another with peak reference demand and another for unloading. Snapshots every 10minutes of the simulation are presented to demonstrate the evolution of traffic flow results and emissions estimations.

It should be noted that the present study provides results that are bounded by the following assumptions:

- The accuracy of the three models used to simulate the different vehicle types, CVs, AVs, and CAVs and the MFC for simulation of vehicle dynamics.

- The accuracy of the two models used to provide $CO_2$ emissions estimates.

- The absence of realistic distribution for the vehicle specifications (mixed vehicle fleet with electric, ICE, hybrid vehicles).

To the authors' knowledge, this is the first study that covers the topic from multiple dimensions (automation, connectivity, vehicle dynamics and driving style) on a large highway network.

Results-wise, among the four different basic scenarios, AVs have the highest emissions. CAVs utilize more the capacity of the network, and therefore during peak hours, they generate more emissions. However, looking at the total values, the differences are not considered significant, probably due to the fixed traffic demand. Furthermore, the car-following models without explicit definition of the vehicle dynamics seem to overestimate instantaneous accelerations, which leads to fast network unloading. The use of MFC describes better the dynamics of the speed, flow and emissions as evolve between consequtive 10-minute intervals during the entire 3-hour simulations. Finally, the instantaneous computation of emissions gives a much more detailed picture regarding emissions estimations. In addition to this, the main conclusions of this work can be summarised as follows:

- Traditional car-following models that aggregate vehicle dynamics, provide overestimated instantaneous acceleration values for certain speeds, which are unrealistic and they artificially unload the transport network at a faster than usual rate.

- AVs have the poorest performance in terms of average speed and flow, and generate the highest emissions values per kilometer.

- CAVs increase the capacity of the network and therefore during peak hours, they generate more emissions in absolute values.

- In total, the differences in emissions per kilometre driven between CVs, AVs, CAVs and CAVs with increased demand do not exceed 6% and thus they are not considered significant.

- The incorporation of the vehicle dynamics in the traditional car-following models using the MFC model depicts better the dynamics of the speed, flow and emissions while progressing between 10-minute periods during the complete 3-hour simulations.

Authors intend to engage in a thorougher study, using variation in the penetration rate of different vehicle types and a realistic vehicle fleet for the simulation of vehicle dynamics. Finally an important topic for future research will be how increasing penetration rates of electrification will impact traffic and emissions.